\documentclass[conference, a4paper]{IEEEtran}
\usepackage{amssymb,amsmath}
\usepackage{amsthm}
\usepackage[]{algorithm2e}
\usepackage[dvips]{graphicx}
\usepackage{color}
\newcounter{proposition}

\newcommand{\nothing}[1]{}

\newcommand{\beq}[1]{\begin{equation}\label{#1}}
	\newcommand{\eeq}{\end{equation}}

\newcommand{\bmu}[1]{\begin{multline}\label{#1}}
	\newcommand{\emu}{\end{multline}}

\newcommand{\eq}{\triangleq}

\renewcommand{\emptyset}{\varnothing}

\renewcommand{\varlimsup}{\mathop{\overline{\lim}}\limits}

\newcommand{\x}{{\textbf{\textit{x}}}}

\newcommand{\e}{{\bf e}}

\newcommand{\F}{\mathcal{F}}

\renewcommand{\S}{{\cal S}}

\renewcommand{\L}{{\bf L}}

\renewcommand{\l}{\ell}

\renewcommand{\S}{{\mathcal{S}}}
\renewcommand{\L}{{\mathcal{L}}}

\newcommand{\s}{{ {s}}}

\renewcommand{\chi}{\upsilon}
\renewcommand{\l}{{ {L}}}

\renewcommand{\le}{\leqslant}
\renewcommand{\ge}{\geqslant}

\renewcommand{\l}{\ell}
\renewcommand{\epsilon}{\varepsilon}
\title{ On Multistage Learning a Hidden Hypergraph}
\author{\IEEEauthorblockN{ A. G. D'yachkov, I.V. Vorobyev, N.A. Polyanskii and V.Yu. Shchukin}
	\IEEEauthorblockA{Lomonosov Moscow State University,
		Moscow, Russia\\
		Email: agd-msu@yandex.ru,\quad vorobyev.i.v@yandex.ru,\quad nikitapolyansky@gmail.com,\quad
		vpike@mail.ru}}
\begin{document}
	\maketitle
	\begin{abstract}
		Learning a hidden hypergraph is a natural generalization of the classical group testing problem that consists in detecting unknown hypergraph $H_{un}=H(V,E)$ by carrying out edge-detecting tests.
		In the given paper we focus our attention only on a specific family $\F(t,s,\l)$  of localized hypergraphs for which  the total number of vertices $|V| = t$, the number of edges $|E|\le s$, $s\ll t$, and the cardinality of any edge $|e|\le\l$, $\l\ll t$. Our goal is to identify all edges of $H_{un}\in \F(t,s,\l)$ by
		using the minimal number of tests.  We develop  an adaptive algorithm that matches the information theory bound,
		i.e., the total number of tests of the algorithm in the worst case is at most $s\l\log_2 t(1+o(1))$. We  also discuss a probabilistic generalization of the problem.
	\end{abstract}	
	\section{Introduction}
	\subsection{Notations and Definitions}
	Let
	$|A|$ denote the size of a set $A$,  $\eq$ denotes the equality by definition
	and $[N]\eq\{1,2,\dots,N\}$ is the set of integers from~$1$ to~$N$.
	A \textit{hypergraph} is  the  pair $H\eq H(V,E)$ such that $E\subset 2^{V}\setminus \varnothing$,
	where $V$ is the set of vertices and 
	$$
	E\eq\left\{(\e_1,\dots \e_s)\,:\, \e_i\subset V, \; i\in[s]\right\}
	$$
	is an $s$-set of edges.  A set $S\subset V$ is called an \textit{independent} set of $H$
	if it does not contain  entire edges of~$H$. We denote by $\dim(H)$ the  
	cardinality of the largest edge, i.e., $\dim(H)=\max\limits_{i\in[s]}|\e_i|$.
	\subsection{Statement of the problem}
	The problem of  learning a hidden hypergraph is described \cite{ad06} as follows. Suppose there is an unknown (hidden) hypergraph $H_{un} = H(V,E)$ whose edges are not known to us, but we know that the unknown hypergraph $H_{un}$ belongs
	to the known family $\F$ of hypergraphs  having certain   specific structure (e.g, $\F$ consists of all Hamiltonian cycles on $V$).
	Our goal is to identify all edges of $E$ by carrying out the minimal number $N$ of \textit{edge-detecting queries} $Q(S)$, where $S\subseteq V$:  $Q(S)$ = 0 if $S$ is independent of $H_{un}$, and  $Q(S)$ = 1 otherwise.
	
	In the given paper we focus our attention only on the family $\F(t,s,\l)$ of  hypergraphs
	introduced in the abstract, namely:  given integers $s$,$\;\l$ and $t$, such that  $s+\l<t$, the set 
	of vertices   $V\eq[t]$ and
	the family $\F(t,s,\l)$,  consists of all hypergraphs $H(V,E)$  such that  $\dim(H)\leqslant \l$ 
	and $|E|\le s$. Suppose we know that the hypergraph $H_{un}$ belongs to the family $\F(t,s,\l)$.  
	An algorithm is said to be an \textit{$\F(t,s,\l)$-searching} algorithm if it finds $H_{un}$, i.e. there exists only one hypergraph from $\F(t,s,\l)$ that fits all answers to the queries.
	
	One of the most important aspects  of any searching strategy is its adaptiveness. An algorithm is \textit{non-adaptive} if all queries are carried out in parallel. An algorithm is \textit{adaptive} if the later queries may depend on the answers to earlier queries.
	
	By $N^{na}(t,s,\l)$ ($N^{a}(t,s,\l)$) denote the minimal number of queries in an 
	$\F(t,s,\l)$-searching non-adaptive (adaptive) algorithms. Introduce the \textit{asymptotic rate} of $\F(t,s,\l)$-searching non-adaptive algorithms:
	$$
	R^{na}(s,\l) \eq \varlimsup\limits_{t\to\infty}\frac{\log_2 t}{N^{na}(t,s,\l)},
	$$
	In similar way we define the asymptotic rate $R^{a}(s,\l)$ of $\F(t,s,\l)$-searching adaptive algorithms.
	
	The rest of the paper is organized as follows. In Sect.~\ref{prev}, we discuss previously known results and remind the concept of cover-free codes which is close to the subject.  In Sect.~\ref{main}, we present the main result of the paper and provide the deterministic adaptive algorithm that matches the information theory bound. Finally,  in Sect.~\ref{AlmostLearning} we  discuss a probabilistic generalization of the problem of learning a hidden hypergraph. 
	
	\section{Previous Results}\label{prev}
	
	For the particular case $\l=1$, the above definitions were already introduced to describe the model called {\em designing screening experiments}. It is a classical group testing problem. We refer the reader to the monograph  \cite{DH} for a survey on group testing and its applications.  It is quite clear (e.g., see \cite{DH}) that an $\F(t,s,1)$-searching adaptive algorithm can achieve the information theory bound, i.e., $N(t,s,1) = s\log_2 t (1+o(1))$ as $t\to\infty$. Therefore, $R^{a}(s,1) = 1/s$. 
	
	If $\l =2$, then we deal with learning a hidden graph. One important application area for such problem is bioinformatics \cite{bg05}, more specifically, chemical reactions and genome sequencing.   Alon et al. \cite{ab04}, and Alon and Asodi \cite{aa05} give lower and upper bounds on the minimal number of tests for
	non-adaptive searching algorithms for certain families of graphs, such as stars, cliques, matchings. In \cite{bg05}, Boevel et al. study the problem of reconstructing a Hamiltonian cycle. In~\cite{a08}, Angluin et al. give a suboptimal $\F(t,s,2)$-searching adaptive algorithm. More precisely, they prove $R^{a}(s,2)\ge 1/(12s)$.
	
	For the general case of parameters $s$ and $\l$, Abasi et al. have recently provided  \cite{AB14} a suboptimal $\F(t,s,\l)$-searching adaptive algorithm. In particular, from their proofs it follows $R^{a}(s,\l) \ge 1/(2s\l)$. This bound differs up to the constant factor from the information theory upper bound $R^{a}(s,\l) \le 1/(s\l)$. In Sect. \ref{main} we improve the lower bound and provide an $\F(t,s,\l)$-searching adaptive algorithm which is optimal in terms of the asymptotic rate.
	\subsection{Cover-Free Codes}
	A binary $N\times t$-matrix
	\beq{X}
	X=\|x_i(j)\|, \quad x_i(j)=0,1,\;i\in[N],\; j\in[t]
	\eeq
	is called a {\em  code of length $N$  and size $t$}. By $\x_i$ and $\x(j)$ we denote the $i$-th row and the $j$-th column of the code $X$, respectively.
	
	Before we give the well-known definition of cover-free codes, note that any $\F(t,s,\l)$-searching non-adaptive algorithm consisting of $N$ queries can be represented by a binary $N\times t$ matrix $X$ such that each test corresponds to the row, and each vertex stands for the column. We put $x_i(j)=1$ if the $j$-th vertex is included to the $i$-th test; otherwise, $x_i(j)=0$.
	
	\textbf{Definition 1.}~\cite{mp88}.\quad
	A code $X$ is called a {\em cover-free $(s,\ell)$-code}
	(briefly, {\em CF $(s,\ell)$-code}) if for any two non-intersecting sets
	$\S,\,\L\subset[t]$, $|\S|=\s$, $|\L|=\ell$, $\S\cap\L=\varnothing$,
	there exists a row $\x_i$, $i\in [N]$, for which
	\beq{property}
	\begin{aligned}
		&x_i(j)=0 \; \text{for any}\;  j\in\S, 
		\\
		&x_i(k)=1\; \text{for any}\; k\in\L.
	\end{aligned}
	\eeq
	Taking into account the evident symmetry over  $s$  and $\l$, we introduce
	$N_{cf}(t,s,\ell)=N_{cf}(t,\ell,s)$ - the minimal length of
	CF $(s,\ell)$-codes of size  $t$ and define
	the {\em asymptotic rate} of  CF $(s,\ell)$-codes:
	\beq{Rsl}
	R_{cf}(s,\ell)=R_{cf}(\ell,s)\eq \varlimsup_{t\to\infty}\frac{\log_2 t}{N_{cf}(t,s,\l)}.
	%\,=\,
	%\varlimsup_{t\to\infty}\frac{\log_2 t}{N_{cf}(t,\s,\ell)}.
	\eeq
	In  \cite{dv02}, Dyachkov et al. show that any CF $(s,\l)$-code represents a $\F(t,s,\l)$-searching non-adaptive algorithm, while any $\F(t,s,\l)$-searching non-adaptive algorithm  corresponds to both a CF $(s,\l-1)$-code and CF $(s-1,\l)$-code. The best presently known  upper and lower bounds on $R(s,\l)$ of CF $(s,\l)$-codes were presented in~\cite{dv02,l03}.
	If $\l\ge1$ is fixed and $s\to\infty$, then these bounds lead to the following asymptotic equality:
	\beq{limits}
	\frac{(\l+1)^{\l+1}}{2e^{\l-1}}\frac{\log_2 s}{s^{\l+1}}(1+o(1))\ge R^{na}(s,\l) 
	$$
	$$
	\simeq R_{cf}(s,\l) \ge \frac{\l^\l}{e^{\l}}\frac{\log_2 e}{s^{\l+1}}(1+o(1)).
	\eeq
	\section{Adaptive Learning a Hidden Hypergraph}\label{main}
	By a counting argument, the lower bound is true.
	
	\textbf{Theorem 1.}\quad \textit{Any $\F(t,s,\l)$-searching algorithm has at least $s\l \log_2{t}(1+o(1))$ edge-detecting queries. In other words, the rate $R^{a}(s,\l)\le 1/(s\l)$.}
	
	\textbf{Proof of Theorem 1.}
	
	Let $N$ be the minimal number of tests in the worst case among all adaptive $\F(t,s,\l)$-searching algorithms. Then we have
	$$
	2^N\ge|\F(t,s,\l)|.
	$$
	This inequity  along with the asymptotic equality 
	$$
	|\F(t,s,\l)|=\frac{t^{s\l}}{(\l!)^s s!}(1+o(1)), \quad t\to\infty,
	$$
	leads to 
	$$
	\frac{\log_2 t}{N}\le \frac{1}{s\l} + o(1), \quad t\to\infty,
	$$
	Therefore, we complete the proof. \qed
	
	The key result of this paper is given as follows.	
	
	\textbf{Theorem 2.}\quad \textit{There exists an adaptive $\F(t,s,\l)$-searching algorithm which has at most $s\l \log_2{t}(1+o(1))$ edge-detecting queries. In other words, the rate $R^{a}(s,\l) = 1/(s\l)$.}
	
	In order to prove Theorem 2 we provide a deterministic $\F(t,s,\l)$-searching adaptive algorithm.
	
	\textbf{Proof of Theorem 2.}
	
	Let $H_{un} = H(V,E)$ be a hidden hypergraph from the family $\F(t,s,\l)$. We will call a vertex $v\in V$ \textit{active}, if there exists at least one edge $\e\in E$ such that $v\in \e$. By $F$, $F\subset V$, denote the set of already found active vertices. By $E'$, $E'\subset E$, denote the set of already found edges of $H_{un}$, i.e., if $\e\in E$ and $\e\subset F$, then $\e\in E'$. Note that a pair $(V,E')$ can be viewed as a partial hypergraph of $H_{un}$. Let $S$, $S\subset V$, be a query we deal with. Before we start the proposed algorithm, we set  $F=\emptyset$, $E'=\emptyset$ and $S=V$.
	
	Firstly we describe an algorithm depicted as Alg.~\ref{AlgoSearchVertex} which allows us to  to find a new active vertex $v$, i.e., $v\not\in F$. The input of the algorithm are set $F$ and a query $S$ which contain at least one edge $\e\in E$ and $\e\not\in E'$. We set $S' = S\setminus F$ and $S'' = S\setminus S'$. At each further step we guarantee that $S'$ contains a new active vertex. While $|S'|>1$, we run the following procedure.  Split up  $S'$ into two equal sized subsets $S_1$ and $S_2$, i.e., $S' = S_1\sqcup S_2$, $|S_1| =\lceil|S'|/2\rceil$ and $|S_2| = \lfloor|S'|/2\rfloor$. Then we carry out a query $S_1\sqcup S''$. If $Q(S_1\sqcup S'') = 1$ then it means that $S_1$ contains at least one new active vertex,  since from the previous steps of the procedure we have $Q(S'')=0$. Therefore we set $S' = S_1$ and repeat the procedure. If $Q(S_1\sqcup S'') = 0$ then at least one new active vertex must lie in $S_2$, since from the previous steps we also have $Q(S_1\sqcup S_2\sqcup S'') = 1$. Thus we set $S' = S_2$, $S'' = S_1\sqcup S''$ and repeat the procedure. At final ($|S'|=1$) we know that the unique vertex $v$ of $S'$ is an active vertex of $H_{un}$ and $v\not\in F$. Notice that Alg.~\ref{AlgoSearchVertex} can be seen  as a variation of the binary vertex search.
	
	Secondly we provide an algorithm depicted as Alg.~\ref{AlgoSearchEdges} which allows us to to find all edges $E'$ composed on already found active vertices $F$. The only input of the algorithm is the set $F$. After we find a new active vertex $v$ we can update the set $E'$ by searching edges containing $v$. But since $|F|\le s\l \ll t$ we can set $E'=\emptyset$ and run the following procedure over all $S$ such that $S\subset F$ and $|S|\le \l$. If there is no edge $\e\in E'$ such that $\e\subset S$, then carry out a query $S$. If $Q(S)=1$ then we delete all edges $\e\in E'$ such that $S\subset \e$ and add edge $\e=S$ to $E'$. Note that Alg.~\ref{AlgoSearchEdges} represents an exhaustive  search of edges.
	
	Thirdly we present an algorithm depicted as Alg.~\ref{AlgoSearchQuery} which allows us to to find a query $S$ such that $S$ contains at least one edge $\e\in E$ and $\e\not\in E'$ (as a consequence $S$ contains at least one active vertex $v\not\in F$). The only input of the algorithm is the set $E'$. We initialize $A$ as vertices included to at least one edge $\e\in E'$, set $B=V\setminus A$ and $S=\emptyset$. One can see that $|A|\le s \l\ll t$. Then we run the following procedure for all sets $C\subset V$ such that $B\subset C$, and $\nexists\, \e\in  E'$, $\e\subset C$. If $Q(C) = 1$ then we set $S=C$ and exit the procedure. If $Q(C) = 0$ we proceed to the next query $C$. If we finish the procedure and have $S=\emptyset$, then it means we have found all edges of $H_{un}$, i.e., $E'=E$. Note that Alg.~\ref{AlgoSearchQuery} is an exhaustive query search.
	
	We give a full description of the proposed $\F(t,s,\l)$-searching algorithm by Alg.~\ref{AlgoMain}, and this algorithm is based on Alg.~\ref{AlgoSearchVertex},  \ref{AlgoSearchEdges} and \ref{AlgoSearchQuery}.  We set $F=\emptyset$, $E'=\emptyset$ and $S=V$. While  Alg.~\ref{AlgoSearchEdges} gives a query $S\neq \emptyset$, we run the following procedure. With the help of Alg.~\ref{AlgoSearchVertex} we find a new active vertex $v$ and add it to $F$. Then we use Alg.~\ref{AlgoSearchEdges} to update the set of edges $E'$ composed on already found active vertices $F$. After that we run Alg.~\ref{AlgoSearchQuery} to find an appropriate query $S$, which then will be used in Alg.~\ref{AlgoSearchVertex}.
	
	Now we upper bound the number of tests of Alg.~\ref{AlgoMain} in the worst case. Let $|V|=t$. It is easy to check that Alg.~\ref{AlgoSearchVertex} uses at most $\lceil \log_2 |S|\rceil\le \lceil \log_2 t\rceil$ tests. One can see that the number of active vertices in the hidden hypergraph $H_{un}\in\F(t,s,\l)$ is at most $s\l$. Alg.~\ref{AlgoSearchEdges} uses at most $F_{1}(s,\l)$ tests, while Alg.~\ref{AlgoSearchQuery} uses at most $F_{2}(s,\l)$ tests, where the functions $F_{1}$  and $F_{2}$ do not depend on $t$. We can upper bound the number of cycles in Alg.~\ref{AlgoMain} by the number of active vertices. Therefore, the total number of tests for the given algorithm does not exceed $s\l(\log_2 t + F_{1}(s,\l) + F_{2}(s,\l) + 1)$.\qed
	\begin{algorithm}[h]\label{AlgoMain}
		\KwData{set of vertices $V$ of  $H_{un}\in \F(t,s,\l)$}
		\KwResult{set of edges $E$ of $H_{un}$}
		initialization $E':=\emptyset$; $F: =\emptyset$; $S: =V$\;
		\While{$S\neq\emptyset$}{
			perform  Alg.~\ref{AlgoSearchVertex}, find  $v\not\in F$ and  $F: = F \sqcup v$\;
			perform  Alg.~\ref{AlgoSearchEdges}, and find subset of edges $E'$\;
			perform  Alg.~\ref{AlgoSearchQuery}, and find query $S$\;
		}
		\caption{Learning a hidden hypergraph}
	\end{algorithm}
	\begin{algorithm}[h]\label{AlgoSearchVertex}
		\KwData{query $S\subseteq V$, $Q(S) = 1$, and set $F\subset V$}
		\KwResult{vertex $v\in V$, $v\not\in F$, and $\exists\, \e\in E$, $v\in \e$}
		initialization $S': = S\setminus F$;  $S'' := S\setminus S'$\;
		\While{$|S'| > 1$}{
			split in half $S'$: $S' = S_1\sqcup S_2$\;
			\eIf{$Q(S_1\sqcup S'') = 1$} 
			{
				$S': = S_1$\;
			}
			{
				$S':= S_2$, $S'':=S''\sqcup S_1$\;
			}
		}
		\caption{Searching a new active vertex}
	\end{algorithm}
	\begin{algorithm}[h]\label{AlgoSearchEdges}
		\KwData{set $F\subset V$}
		\KwResult{subset of edges $E'\subset E$}
		initialization  $E': = \emptyset$\; 
		\For{$\forall S\subset F$: $1\le|S|\le\l$}{
			\If{$\nexists\, \e\in E':$ $\e\subset S$}{
				\If{Q(S) = 1
				}{
				\For{$\forall \e\in E':$ $S\subset \e$}{
					delete $\e$ from $E'$\;
				}
				add edge $\e = S$ to $E'$\;
			}  
		}
	} 
	\caption{Searching edges}
\end{algorithm}
\begin{algorithm}[h]\label{AlgoSearchQuery}
	\KwData{subset of edges $E'\subset E$}
	\KwResult{query $S$}
	initialization $A := \{v: \; v\in \e\in E'\}$; $B: = V\setminus A$; $S: = \emptyset$\;
	\For{$\forall C\subset V$: $B\subset C$ and $\nexists\, \e\in  E'$, $\e\subset C$ }{
		\If{Q(C) = 1} 
		{
			$S := C$ and
			break ``for loop''\;
		} 
	}
	\caption{Searching a query}
\end{algorithm}
\section{Concept of ``Almost'' Learning a Hidden Hypergraph}\label{AlmostLearning}
Remind that there are two natural types of algorithms, namely non-adaptive and adaptive. A compromise between these two types is usually called a \textit{multistage} (or \textit{multiple round}) algorithm. In the given section we  will limit ourselves to the consideration of only so called \textit{two-stage searching procedures} (see, e.g., \cite{BGV}). It means we can adapt the tests of the second round of testing only one time after we receive the answers to the queries of the first round.

Now we consider a relaxation of the problem of learning a hidden hypergraph.  Suppose we let an algorithm identify \textit{almost all} hypergraphs  from the family $\F(t,s,\l)$. More formally, if there exist a subfamily $\F'(t,s,\l)\subset \F(t,s,\l)$ of cardinality at least $(1-\epsilon)|\F(t,s,\l)|$ such that for any $H_{un}\in \F'(t,s,\l)$ the algorithm finds $H_{un}$, then
we say that the algorithm is \textit{$\F(t,s,\l)$-searching with probability $(1-\epsilon)$}. 

By $N^{na}(t,s,\l,\epsilon)$ ($N^{a}(t,s,\l,\epsilon)$, $N^{2st}(t,s,\l,\epsilon)$) denote the minimal number of queries in a $\F(t,s,\l)$-searching non-adaptive (adaptive, two-stage) algorithm with probability $(1-\epsilon)$. Introduce the \textit{capacity} of non-adaptive $\F(t,s,\l)$-searching algorithms
$$
C^{na}(s,\l) \eq \varlimsup\limits_{\substack{\epsilon\to 0
		\\ t\to\infty }
}\frac{\log_2 t}{N^{na}(t,s,\l,\epsilon)}.
$$
In similar way we define  the capacities $C^{a}(s,\l)$ and $C^{2st}(s,\l)$ of adaptive $\F(t,s,\l)$-searching algorithms and two-stage $\F(t,s,\l)$-searching algorithms, respectively. 

One can easily check that Theorem $1$ is true for ``almost'' concept as well. From definitions it follows
$$
C^{na}(s,\l) \le C^{2st}(s,\l) \le  C^{a}(s,\l) = \frac{1}{s\l},
$$
where the right-hand side equality holds in virtue of Theorem~2.

For the case $\l=1$ the definition of the capacity was considered in many papers. We refer the reader to the classic result \cite{m78} in model of designing screening experiments. Malyutov proved that $C^{na}(s,1) = 1/s$.

We conjecture that $C^{na}(s,\l)=1/(s\l)$. The following theorem reinforces the hypothesis.

\textbf{Theorem 3.}\quad \textit{The capacity of two-stage $\F(t,s,\l)$-searching algorithms $C^{2st}(s,\l) = 1/(s\l)$.}

We prove Theorem 3 using the probabilistic method and the result established in \cite{m78} for the case $\l=1$.

\textbf{Proof of Theorem 3.}

Firstly we note that the subfamily $\F'(t,s,\l)\subset \F(t,s,\l)$  which consists of $s$ pairwise non-intersecting edges of size $\l$, i.e.,
$$
\F'(t,s,\l)= \left\{
\begin{aligned}
H\in\F(t,s,\l):\; H=(\e_1,\dots,\e_s),\\ |\e_i| = \l,\;\e_i\cap \e_j = \varnothing\text{ for any }i\neq j
\end{aligned}
\right\},
$$ 
has cardinality $|\F(t,s,\l)|(1+o(1))$ as $t\to\infty$. Thus, for any $\epsilon>0$ and for sufficiently large $t$ it is enough to prove the existence of two-stage $\F'(t,s,\l)$-searching algorithm with probability $(1-\epsilon)$. 	

Now we show that there exists a binary matrix (each test corresponds to the row, and each vertex stands for the column) corresponding to the first stage of group testing such that after carrying out tests of the first stage for almost all hypergraphs from $\F'(t,s,\l)$ we can find a \textit{good} partition of vertices into $s$ disjoint sets: $V_1\sqcup V_2\dots \sqcup V_s = V = [t]$, such that $\e_1\in V_1,\,\e_2\in V_2,\ldots, \e_s\in V_s$.
Define the ensemble $E(N,t,s)$ of $s$-ary $(N\times t)$-matrices $X=||x_i(j)||$,
where each entry $\x_i(j)$
is chosen independently and equiprobably from the set $\{1,2,\ldots, s\}$. Each $s$-ary symbol $x$ in $X$ is then replaced by the binary column of length $s$, which has only one $1$ at $x$-th position. In other words, the $s$-ary $(N\times t)$-matrix $X$ is replaced by the binary $(sN\times t)$-matrix $X_1$, and each $s$-ary row of $X$ is replaced by the binary $(s\times n)$-layer of $X_1$ such that each column of the layer contains only one $1$. 
Define the event $A(s,\l)$: ``given a hypergraph $H\in \F'(t,s,\l)$, there exists a layer in $X_1$ such that the answers to all $s$ edge-detecting queries in the layer are $1$'s''. If this condition holds, then we are presented with a good partition into disjoint sets: $V_1\sqcup V_2\dots \sqcup V_s = V = [t]$, such that $\e_1\in V_1,\,\e_2\in V_2,\ldots, \e_s\in V_s$. Denote the cardinality of $V_i$ by $t_i=|V_i|$. Now estimate the probability of the opposite event to $A(s,\l)$:
$$
P\left(\overline{A(s,\l)}\right)=\left(1-\frac{s!}{s^{s\l}}\right)^{N}.
$$
Let $N\to \infty$ and $N=o(\log_2t)$. Then for any $\epsilon>0$ there exists $N(\epsilon)$ such that for any $N\ge N(\epsilon)$ the probability $P\left(\overline{A(s,\l)}\right)\le \epsilon$. It means that for any $\epsilon>0$, and for sufficiently large $t$ and for $N=o(\log_2t)$ (the number of tests of the first stage is negligible) there exists a binary $(sN\times t)$-matrix $X_1$ which can be used in the first stage of the algorithm, and for at least $(1-\epsilon)|\F'(t,s,\l)|$ hypergraphs from $\F'(t,s,\l)$ there exist $s$ edge-detecting queries, answers of which are $1$'s, and supports of these queries are pairwise non-intersecting. For matrix $X$ we will call all such hypergraphs \textit{good}. For any binary $(N\times t)$ matrix $X$ denote the set of good hypergraphs by $G(X)$, $G(X)\subset \F'(t,s,\l)$. Notice that we have proved $|G(X_1)|\ge (1-\epsilon)|\F'(t,s,\l)|$.

Now we reformulate the statement derived in~\cite{m78}.

\textbf{Lemma 1.} \quad \textit{The capacity of non-adaptive $\F(t,\l,1)$-searching algorithms  $C^{na}(\l,1) = 1/\l$.}

Notice that if we are given with a code $X$ representing a non-adaptive $\F(t,\l,1)$-searching algorithm with probability $(1-\epsilon)$ then we can replace each entry of $X$ by the following rule: $0\to 1$, $1\to 0$, and get the code $Y$, which represents a non-adaptive $\F(t, 1,\l)$-searching algorithm with probability $(1-\epsilon)$.

Roughly speaking, for any good hypergraph from $G(X_1)$ it will be sufficient to apply $s$ non-adaptive $\F(t_i, 1,\l)$-searching algorithms with probabilities $(1-\epsilon)$ in parallel at the second stage of our strategy in order to find $s$ edges in each $V_i$, $t_i=|V_i|$.

More formally, let $E(N, t, X)$ be the ensemble of binary $(N\times t)$ codes that consists of all possible permutations of columns of a code $X$  representing a non-adaptive $\F(t,1,\l)$-searching algorithm with probability $(1-\epsilon)$, and each copy of $X$ is chosen equiprobably with probability $1/t!$. Let $Y$ be a random matrix of $E(N, t, X)$.  For a good hypergraph $H\in G(X_1)$ let we be given with an appropriate partition into disjoint sets $V_1\sqcup V_2\dots \sqcup V_s = V = [t]$ such that $\e_1\in V_1,\,\e_2\in V_2,\dots \e_s\in V_s$.  At the second stage of our strategy for vertices $V_1$ we will apply $N$ tests of $Y$ without using vertices $V\setminus V_1$, for vertices $V_2$ we will apply $N$ tests of $Y$ without using vertices $V\setminus V_2$ and so on. Define the event $B(s,\l)$: ``all edges of $H$ can be found applying s$N$ tests''. Estimate the probability of the opposite event to $B(s,\l)$:
$$
P(\overline{B(s,\l)})\le s\epsilon.
$$
It means that there exists such a code $X_2$ which is a copy (obtained by a column permutation) of $X$ such that for $(1- s\epsilon)\cdot |G(X_1)|$  good hypergraphs we find all edges after the second stage of our strategy.

Finally, for any $\epsilon>0$ there exists an ascending sequence of $t$ such that for $(1-\epsilon)|\F(t,s,\l)|$ hypergraphs there exists a code $X_1$, which can be used for the first stage  of $\F(t,s,\l)$-searching algorithm with probability $(1-\epsilon)$, and a code $X_2$, the modification of which can be applied for the second stage of the strategy, such that the total number of tests is sufficiently determined by only queries of the code $X_2$ and is equal to $s\l\log_2t(1+o(1))$. \qed

\end{document}